# Formation of ultracold RbCs molecules by photoassociation

N. Bouloufa-Maafa[a,c], M. Aymar[a], O. Dulieu[a] and C. Gabbanini[b*]

[a] Laboratoire Aimé Cotton, CNRS, Bât. 505, Univ. Paris-Sud,

F-91405 Orsay Cedex, France

[b] Istituto Nazionale di Ottica, INO-CNR, U.O.S. Pisa ''Adriano Gozzini'',

Via G. Moruzzi 1, 56124 Pisa, Italy

[c] Université Cergy-Pontoise, 95000 Cergy-Pontoise, France

e-mail: carlo.gabbanini@ino.it

**Abstract**

The formation of ultracold metastable RbCs molecules is observed in a double species magneto-optical trap through photoassociation below the $^{85}$Rb($5S_{1/2}$)+$^{133}$Cs($6P_{3/2}$) dissociation limit followed by spontaneous emission. The molecules are detected by resonance enhanced two-photon ionization. Using accurate quantum chemistry calculations of the potential energy curves and transition dipole moment, we interpret the observed photoassociation process as occurring at short internuclear distance, in contrast with most previous cold atom photoassociation studies. The vibrational levels excited by photoassociation belong to the $5^{th}$ $0^+$ or the $4^{th}$ $0^-$ electronic states correlated to the Rb($5P_{1/2,3/2}$)+Cs($6S_{1/2}$) dissociation limit. The computed vibrational distribution of the produced molecules shows that they are stabilized in deeply bound vibrational states of the lowest triplet state. We also predict that a noticeable fraction of molecules is produced in the lowest level of the electronic ground state.

## 1. Introduction

In the last fifteen years the field of cold molecules reached important advances and demonstrated new applications in many research fields. Among them we can cite fundamental tests in physics, molecular clocks, molecular spectroscopy, ultracold reactions, controlled photochemistry and quantum computation. Recent reviews on the status of the subject and on possible perspectives can be found in Refs.[1–5]. In particular cold polar molecules, *i.e.* that possess a permanent electric dipole moment (PEDM) in their body-fixed frame, can be utilized for quantum information. Polar molecules can be manipulated by external electric fields, allowing for the control of elementary chemical reactions at very low temperatures. Heteronuclear alkali molecules in the lowest vibrational level of their ground state $X^1\Sigma^+$ possess a PEDM ranging between 0.5 to 5.5 Debye [6], giving rise to long range anisotropic dipole–dipole interactions in the presence of an external polarizing electric field.

In this paper, we produce ultracold RbCs molecules utilizing the process of photoassociation (PA) followed by radiative stabilization [7, 8]. PA bounds together atoms of two different alkali species into excited state molecules that can decay by spontaneous emission into (meta-) stable molecules. The process produces molecules that are generally distributed in high-lying rovibrational levels of the ground singlet or lowest metastable triplet state [9, 10], while evidence for the spontaneous creation of deeply-bound ground state polar bi-alkali molecules such as KRb [11], LiCs [12], and NaCs [13, 14] has been reported. Molecules in high-lying vibrational levels have a weakly polar character but represents a starting point for a further manipulation that can transfer them into the lowest vibrational level *v=0*. This can be achieved by an incoherent optical process [15] which has been actually the first demonstration of the controlled transfer of the vibrational population of ultracold molecules in the *v=0* level of the RbCs ground state. Later, the cooling of the vibrational degree of freedom utilizing a properly shaped laser [16] has been demonstrated on ultracold $Cs_2$ molecules, and only recently on a polar species, NaCs [17]. The coherent transfer of the population of a single highly-excited level of $^{41}K^{87}Rb$ ground state molecules produced by PA down to *v=0* using the stimulated Raman adiabatic passage process (STIRAP) [18] has been also achieved. Note that the latter method has been proven very successful when starting from much colder atomic ensembles close to quantum degeneracy: ultracold molecules in their absolute ground state can be obtained by transferring the population of the long-range state initially created by magnetoassociation of both homonuclear [19-21] and heteronuclear species [22, 23]. All these methods require the knowledge of the rovibrational distribution of the created cold molecules, as well as detailed insights on molecular spectroscopy and dynamics.

In a previous paper [24] (hereafter referred to as paper I) we reported on a new PA path leading to the formation of ultracold RbCs molecules in a double species magneto-optical trap (MOT), and proposed a preliminary interpretation relying on PA at short internuclear distance. Here we investigate this process in more details and we predict that ultracold RbCs molecules can be formed in deeply-bound vibrational levels (including *v=0*) of the lowest electronic triplet state and of the ground state. In Section 2 we recall the main features of our experimental setup which lead to the observation of ultracold RbCs molecule formation (Section 3). The required computations of the RbCs molecular structure including potential curves, transition dipole moments and a model for spin-orbit coupling are presented in Section 4, allowing the interpretation of the experimental results through the computation of PA probabilities and of vibrational distributions by spontaneous emission (Section 5). Prospects for the experimental characterization of these distributions as well as for possible implementation of new path for vibrational population down to the lowest molecular level are discussed in the conclusion.

## 2. Experimental setup

The experimental apparatus is sketched in Fig.1. A double species MOT is produced inside a UHV metal chamber, loading both species from vapour; for Cs the background gas pressure is controlled by a valve from the vapour in equilibrium with a metal sample, while for Rb it is emitted by running current through a metal dispenser. The MOT is produced by two DFB diode lasers, frequency tuned about two linewidths below the standard cooling transitions ($6S_{1/2}(F = 4)$ - $6P_{3/2}(F = 5)$ and $5S_{1/2}(F = 3)$ - $5P_{3/2}(F = 4)$ for Cs and $^{85}$Rb respectively), together with two additional diode lasers, tuned on transitions from the other ground hyperfine levels of the two atoms, as repumpers. All lasers are frequency-locked in different cells using saturated absorption spectroscopy. The repumper beams can be almost extinguished passing through acousto-optic modulators. Along the vertical direction, which is also the axis of the quadrupole magnetic field, the lasers for Cs and Rb have superposed paths while on the horizontal plane there are separated retroreflected beams for the two species. This geometry permits to independently control the alignment for each species and therefore to maximize the overlap of the two cooled clouds, that is monitored by two CCD cameras along orthogonal axis. About $10^7$ atoms of both Cs and $^{85}$Rb atoms are trapped with densities of a few $10^{10}$ cm$^{-3}$. The temperature of the sample has not been directly measured in this experiment, but should be between 100 and 200 µK as in other similar setups.

On the trapped sample a PA laser, consisting of a tapered amplifier injected by a DFB diode laser, is focused. The PA laser can be tuned by about thirty wavenumbers around the Cs D2 line by changing the temperature of the DFB diode laser. The frequency scan of the PA laser is monitored by a Fabry-Pérot interferometer that can be used also to frequency lock the DFB laser, while a wavemeter measures the absolute frequency.

A pulsed dye laser, with energy of about 1 mJ, pumped by the second harmonics of a Nd:YAG laser (20 Hz repetition rate and 7 ns time width) is softly focused on the cold cloud in order to ionize the produced molecules. Molecules are ionized by resonant enhanced two-photon ionization (*RE2PI*). The dye laser works with LDS 698 dye covering the frequency range from 13800 to 14800 cm$^{-1}$. Before the arrival of the laser pulse, the repumping lasers of the double MOT are almost extinguished for about 10 ms, producing a temporal dark SPOT. This causes a decrease of the collisional losses (both intra-species and inter-species) and an initial increase of the atomic densities. The atomic and molecular ions created by the laser pulse are repelled by a grid and detected by a microchannel plate after by time-of-flight selection.

### 3. Experimental results

Although PA below the Rb($5S_{1/2}$)+Cs($6P_{3/2}$) dissociation limit induces predissociation [14], we found that for a few particular frequencies stable RbCs molecules are created, as demonstrated by the ion signal obtained by *RE2PI* at a frequency of 14169.7 cm$^{-1}$ shown in Fig. 2a. The three PA lines recorded at frequencies

11724.044(15), 11724.072(15), and 11724.127(15) cm$^{-1}$ correspond to the $J = 0, 1, 2$ rotational levels of a single vibrational level of a molecular state excited by PA (Fig. 2b). As the observed rotational progression starts with $J=0$, the excited electronic state should have a value $\Omega=0$ of the projection of the total electronic angular momentum on the molecular axis. Moreover the lines do not exhibit any significant hyperfine structure, which is also in favour of a $\Omega=0$ state.

The corresponding rotational constant is $B = 0.0138(7)$ cm$^{-1}$, a large value indicating that the PA process acts at short range. The $J = 1$ line displays the Stark effect, consisting in a splitting and shift of the line, due to the static electric field used to extract the ions from the trapping region and direct them to the microchannel plate. Our electric field control prevented us to precisely quantify this effect. Over the whole frequency region where the PA laser can be tuned, no other lines have been observed, even at small detuning from the dissociation limit, in agreement with the hypothesis of a strong predissociation.

The *RE2PI* spectra have been recorded by scanning the pulsed dye laser, with the PA laser frequency locked on a transmission peak of a stable Fabry-Pérot interferometer, tuned in correspondence of one of the observed PA transitions (the $J=1$ line in the present case). A dense spectrum of the whole investigated region is shown in Fig.2c. A strong detection band extends between 14000 and 14300 cm$^{-1}$, while weaker transitions are also observed between 14400 and 14600 cm$^{-1}$.

4. **The molecular states in the region of the Rb(5S)+Cs(6P$_{3/2}$) limit.**

The observation of an isolated PA line with large rotational constant hints the occurrence of a short-range process, where a vibrational level of a molecular potential well located near the Rb($5S_{1/2}$) + Cs($6P_{3/2}$) dissociation limit is directly populated. The formation of heteronuclear alkali molecules through a short-range PA process was already observed for LiCs molecules [12]. In paper I, we based our interpretation on the analysis of the computed RbCs molecular potentials including spin–orbit interaction [25]: there are several potential wells correlated to the Rb($5P_{1/2,3/2}$) + Cs($6S_{1/2}$) dissociation limits with a minimum located below the Rb($5S_{1/2}$) + Cs($6P_{3/2}$) energy, hence offering the possibility for short-range PA of their levels.

In order to compute transition probabilities for PA and cold molecule formation, we calculated RbCs potential curves according to the method of Ref. [6] also used by the authors of Ref [25]. Briefly, the RbCs molecule is modelled as two-valence-electrons moving in the field of a $Rb^+$ and a $Cs^+$ polarizable cores separated by a distance R. The ionic cores are represented with effective core potentials (ECP) [26] including core-polarization potentials (CPP) [27] with parameters adjusted on experimental atomic energies [28]. A full configuration interaction within a two-electron configuration space generated by a large Gaussian basis set yields potential energy curves in Hund's case *a*. The empirical short-range correction reported in Ref.[29] is added to all PECs. At this step our results are obviously very similar to those of Ref. [25], with small differences arising from the choice of the ECP and CPP parameters [28] which are not significant for the present purpose. As in Ref. [23] we replaced (Figure 3b) the computed X and a curves by their experimental determination of Refs. [30, 31]. The corresponding transition dipole moments (TDM) from the ground state X $^1\Sigma^+$ and the lowest triplet state a $^3\Sigma^+$ (hereafter referred to as the X and a states, respectively) are drawn in Figure 3c [1].

In order to use the TDMs above we need to model the spin-orbit couplings between Hund's case a states, which are not available in Ref.[25]. We built electronic states including spin-orbit interaction using a perturbative approach restricted to the $(3)^1\Sigma^+$, $(3)^3\Sigma^+$, $(2)^1\Pi$, $(2)^3\Pi$ states correlated to the Rb($5^2$P)+Cs($6^2$S) asymptote. This results into well-known potential energy Hamiltonian matrices [32] that we recall below for readers' convenience:

$$H(\Omega = 0^{+,-}) = \begin{pmatrix} V((2)^3\Pi) - \Delta E_{fs}/3 & (\sqrt{2}/3)\Delta E_{fs} \\ (\sqrt{2}/3)\Delta E_{fs} & V((3)^{1,3}\Sigma^+) \end{pmatrix} \quad (1)$$

$$H(\Omega = 1) = \begin{pmatrix} V((2)^3\Pi) & \Delta E_{fs}/3 & \Delta E_{fs}/3 \\ \Delta E_{fs}/3 & V((2)^1\Pi) & -\Delta E_{fs}/3 \\ \Delta E_{fs}/3 & -\Delta E_{fs}/3 & V((3)^3\Sigma^+) \end{pmatrix} \quad (2)$$

The matrix for the $\Omega$=2 component reduces to a single element H($\Omega$=2)=V((2)$^3\Pi$)+$\Delta Efs$/3. For simplicity, we used an *R*-independent spin-orbit coupling determined by the atomic spin-orbit splitting of the Rb($5^2$P) level $\Delta E_{fs}$=237.6 cm$^{-1}$. The matrices are diagonalized providing the PECs displayed in Figure 3a. The PECs are labelled according to their dominant Hund's case *a* character at short internuclear distances, with subscripts related to their Hund's case *c* symmetry. They differ from those of Ref.[25] by typically a few tens of wavenumbers, which indicates that the *R*-variation of the spin-orbit coupling, while noticeable, will not modify significantly the proposed interpretation, as long as spectroscopic accuracy is not concerned.

---

[1] The numerical data for potential energy curves and transition dipole moments can be obtained upon request to the authors.

At first glance we see that the spin-orbit components of at least two Hund's case a states dissociating toward Rb($5^2$P)+Cs($6^2$S) have their minimum in the region of the Rb($5S_{1/2}$)+Cs($6P_{3/2}$) limit: the $(2)^3\Pi$ and the $(2)^1\Pi$ states. The $(3)^3\Sigma^+$ state may be a candidate as the computed curves are not accurate enough to predict the energy position of the minimum to better than several tens of wavenumbers. The levels of the $(3)^1\Sigma^+$ state are also accessible but they will not match the interpretation of the present results due to their small rotational constant, as discussed below. For future reference the spin-orbit components of these states at short internuclear distances are indexed using conventional notations as follows: $(5)0^+$, $(4)0^-$, $(6)1$, and $(2)2$ for the $(2)^3\Pi$ state, $(5)1$ for the $(2)^1\Pi$ state, $(5)0^-$ and $(7)1$ for the $(3)^3\Sigma^+$ state, and $(4)0^+$ for the $(3)^1\Sigma^+$ state.

However, such an isolated potential description is by far insufficient to model the system. This is demonstrated by the results reported in Figure 4 for the rotational constant of the vibrational levels of the system, computed as follows. Starting from the Hamiltonian matrix above, the vibrational wave functions of the coupled state problem are obtained with the Mapped Fourier Grid Hamiltonian method extensively detailed elsewhere [33, 34]. The rotational constant $B_v$ is defined by:

$$B_v(\Omega) = \left\langle \Xi_v(\Omega) \left| \frac{1}{2\mu R^2} \right| \Xi_v(\Omega) \right\rangle \qquad (3)$$

where $\Xi_v(\Omega; R)$ is the multi-component (on the electronic states involved in each of the $\Omega=0^+, 0^-, 1$ symmetries) radial wave function labelled by the index $v$ ordering the levels, $\mu$ the reduced mass of the $^{85}$Rb$^{133}$Cs molecule, and the brackets denote the integration on the $R$ coordinate. Formally we write them as:

$$\Xi_v(0^+;R) = \alpha_v^\Pi(R)\,|(2)^3\Pi\rangle + \alpha_v^\Sigma(R)\,|(3)^1\Sigma^+\rangle,$$
$$\Xi_v(0^-;R) = \beta_v^\Pi(R)\,|(2)^3\Pi\rangle + \beta_v^\Sigma(R)\,|(3)^3\Sigma^+\rangle, \qquad (4)$$

$$\Xi_v(1;R) = \gamma_v^\Pi(R)\,|(2)^3\Pi\rangle + \gamma'^{\Pi}_v(R)\,|(2)^1\Pi\rangle + \gamma_v^\Sigma(R)\,|(3)^3\Sigma^+\rangle,$$

where the kets hold for the relevant Hund's case *a* electronic states. Figure 4 strikingly reveals that even at short internuclear distances the Hund's case *a* states are mixed by the spin-orbit coupling: the rotational constants of levels belonging to the $\Omega=0^+$, $0^-$, 1 symmetries clearly deviates in several places from the ones of Hund's case *a* curves. Note however that the lowest trace (closed circles) in that figure relates to levels mainly located within the $(3)\,^1\Sigma^+$ well, with a rotational constant far too small compared to the observed one. Similarly, the uppermost trace (full triangles) correspond to levels located within the $(3)\,^3\Sigma^+$ well with a too large value of $B_v$. Figure 5 focuses on the rotational constants in the region of interest, namely, around the energy of the $Rb(5S_{1/2})+Cs(6P_{3/2})$ limit. In addition we displayed the weights defined in Equation 4 above of the vibrational wavefunctions on the Hund's case *a* states. It confirms that in this region the $0^+$ levels have a strong Hund's case *a* state character (panel b) while the levels of the $\Omega=0^-$, 1 symmetries are strong mixtures of these states (panels d and f). This will be of central importance for the calculations of the next section. Note that levels of each of the three symmetries are possible candidates for the observed line. The computed rotational constants are slightly smaller than the experimental one, probably due to the approximate model we used for the spin-orbit coupling.

## 5. Transition probabilities for PA and for cold molecule formation

The efficiency of the PA process is well described by both the squared overlap integral between the radial wavefunction $\Psi(E,i)$ of initial state *i* of the colliding pair and those of the vibrational levels $\Xi_v(\Omega; R)$:

$$D_v(E, \Omega) = |\langle \Psi(E,i) | \Xi_v(\Omega) \rangle|^2 \tag{5}$$

and by the related squared matrix element of the *R*-dependent TDM $d^{i\text{-}\Omega}(R)$:

$$D_v(E, \Omega) = \left|\langle \Psi(E,i) | d^{i-\Omega} | \Xi_v(\Omega) \rangle\right|^2 \tag{6}$$

The initial wave function $\Psi(E,i)$ is an undetermined mixture of the $X^1\Sigma^+$ and $a^3\Sigma^+$ states, but as we assume that PA proceeds at short distances, we can restrict it to its triplet component which has a turning point (and hence a local maximum) around 9.8 a.u.:

$$\Psi(E,i) = \chi(R;E,i)|a^3\Sigma^+\rangle \tag{7}$$

Indeed at such distances the radial wavefunctions in the X channel oscillates very rapidly and we numerically checked that it yields a negligible overlap with the short-range vibrational wave functions of the PA levels. In the following, the vibrational wave functions of the X and a states are obtained with the MFGH method as above. The continuum wave function $\chi(R; E, a)$ of the a state at a scattering energy $E$ equivalent to a temperature $T=E/k_B=290\mu K$ is taken after adjusting the grid of the MFGH calculation such that the first (unity-normalized) solution above the dissociation threshold possesses this energy.

Under this assumption, we see on Fig. 6 that PA relies on the excitation of the $(2)^3\Pi$ component of the wave functions in a similar amount for all $\Omega=0^+$, $0^-$, 1 symmetries. While having a TDM with the a state of similar magnitude than $(2)^3\Pi$, the $(3)^3\Sigma^+$ state has its minimum at too small distance to be efficiently reached from the a state. To be significant, the magnitude of the quantities reported in Fig. 6 must be rescaled with the energy normalization of the continuum wavefunction and with factors depending on the experiment (intensity and wavelength of the PA laser).

Figure 7 shows the squared overlap integrals of the wavefunctions of selected PA levels (marked with crosses in Fig. 5) in each symmetry with those of the X and a states. These quantities are now representative of the efficiency of this second step of the cold molecule formation process as they involve unity-normalized wavefunctions of bound levels. For illustration purpose, the chosen levels in the $\Omega=0^+$, $0^-$ cases have a main component on the $(2)^3\Pi$ state, while the one for $\Omega=1$ has its main component on the $(2)^1\Pi$ state. In the three upper panels the decay of the $(2)^3\Pi$ radial component of the $\Omega=0^+$, $0^-$,1 levels is represented showing different patterns from one panel to the other: results in panels a and b exhibit the same envelope, confirming that the chosen levels have a strong $(2)^3\Pi$ character only slightly perturbed by the other state. Note that a similar pattern is expected in levels of the $\Omega=2$ curve (identical to the shifted $(2)^3\Pi$ curve) would be excited by PA. In panel c the envelope shows once more that the electronic states are coupled in different ways for each symmetry: this results into strong perturbations in these radial wave functions, compared to what would be the radial wavefunction of an isolated pure $(2)^3\Pi$ level. In particular, the distortion brought to the $(2)^3\Pi$ by the spin-orbit coupling with the $(2)^1\Pi$ state (whose outer turning point is almost aligned with the minimum of the $a^3\Sigma^+$ state) is predicted to create ultracold molecules in the $v=0$ level of the $a^3\Sigma^+$ state. An even more promising situation is observed in Fig. 7d: the distortion of the $(3)^1\Sigma^+$ component by the spin-orbit coupling with the $(2)^3\Pi$ state induces a high formation rate in the $v=0$ levels of the RbCs ground state, which can be related to the almost perfect alignment of the inner turning point of the $(5)0^+$ curve with the minimum of the X curve (see Fig. 3). The chosen $\Omega=1$ level is expected to produce a broad vibrational distribution on the X levels, with a noticeable amount toward the $v=0, 1$ ground state levels as well.

## 6. Concluding remarks and perspectives

In this paper we analyzed the formation mechanism of ultracold RbCs molecules through photoassociation near the $Rb(5S_{1/2})+Cs(6P_{3/2})$ dissociation limit followed by radiative stabilization. The observed photoassociation process involves the excitation at short internuclear distance of a vibrational level of the $(5)0^+$ or the $(4)0^-$ electronic states correlated to the $Rb(5P_{3/2})+Cs(6S_{1/2})$ and $Rb(5P_{1/2})+Cs(6S_{1/2})$ dissociation limit, respectively. The computed vibrational distributions of the RbCs molecules strongly suggest they are created in deeply-bound vibrational states of the lowest triplet state $a^3\Sigma^+$ (for both possible states $(5)0^+$ or $(4)0^-$), and possibly with a noticeable fraction of molecules in the lowest vibrational levels of the $X\ ^1\Sigma^+$ ground state (only if the $(5)0^+$ state is populated). The modelling of the experimental PA spectrum suggests that other PA levels belonging to the $\Omega=1$ symmetry could well be excited close to the observed ones. As the vibrational distribution of the created cold molecules is predicted to be quite different than in the $\Omega=0^{+,-}$ cases, their detection probably relies on the choice of a different frequency of the ionizing laser pulse. The extension of the range of the PA laser detunings down to the $Rb(5S_{1/2})+Cs(6P_{1/2})$ asymptote should also reveal many other PA lines, with the possibility to determine the energy position of the lowest levels of the relevant molecular potential curves.

The interpretation of the *RE2PI* spectrum of Fig. 2c is in progress and will be presented in a future publication. Between 14000 and 14300 cm$^{-1}$, the spectrum presents many intense grouped in three overlapping packets approximately cantered at 14090, 14180, and 14270 cm$^{-1}$. Such a pattern has already been observed in a previous work on Cs$_2$ photoassociation [35], and has been assigned to the *RE2PI* of ultracold Cs$_2$ molecules in their lowest triplet state a$^3\Sigma_u^+$ via the three spin-orbit components $\Omega$=0, 1, 2 of the excited (2)$^3\Pi_g$ correlated to Cs(6$^2$S)+Cs(5$^2$D). It is likely that a similar interpretation still holds in the present case (namely, the ionization of levels of the a $^3\Sigma^+$ levels), as the quantum chemistry calculations predict the presence of the bottom of the (3)$^3\Pi$ potential (correlated to Rb(5$^2$S)+Cs(5$^2$D)) around 14000 cm$^{-1}$ (see for instance Fig.2 of paper I). Another molecular state, the (4) $^3\Sigma^+$ state, is also predicted with a potential well located around 14500 cm$^{-1}$. This could well explain the less intense lines observed in this region, again in a way very similar to the comparable analysis of Cs$_2$ RE2PI spectra reported in Ref. [36].

Future experiments will look for state-selective detection of the expected singlet state molecules. According to the present study, this simple short-range PA process looks promising to produce a sample of polar RbCs molecules, either by accumulating in a trap the molecules directly formed in the lowest level of the electronic ground state or by performing vibrational cooling [16] on the molecular distribution. Furthermore, a very interesting prospect is the possibility to achieve an efficient STIRAP scheme for transferring ultracold RbCs molecules initially created in a Feshbach resonance with dominant a$^3\Sigma^+$ character down to the *v*=0 level of the electronic ground state. This is suggested by Fig. 4a and Fig. 5d: the first step of the STIRAP process could rely on the excitation toward a level of the 0$^+$ state (with a significant (3)$^3\Pi$ character) at a frequency close to the one of the Cs D2 line (852 nm), while the second step would correspond to the transition from the (3)$^1\Sigma^+$ component of the same 0$^+$ level down to $v$=0 of the X state around 15500 cm$^{-1}$ or 645 nm. A more quantitative evaluation of the efficiency of this transition is currently under investigation in our group.

## 7. Acknowledgements

**Figure captions**

Figure 1: Sketch of the experimental apparatus. TA: tapered amplifier, FPI: Fabry-Pérot interferometer, MCP: microchannel plate.

Figure 2: (a) Recording of the ion signal after time-of-flight selection. The ionization laser frequency is fixed at 14169.7 cm$^{-1}$. (b) photoassociation spectrum recorded at the same ionization frequency showing the rotational structure ($J$=0, 1, 2). The origin correspond to the position of the $J$=0 line at 11724.044(15) cm$^{-1}$; (c) Molecular ion spectrum with the PA laser locked on the $J$=1 line at 11724.072(15) cm$^{-1}$. The peaks and dips near 14340 cm$^{-1}$ are due to resonant atomic ionization ($6^2P$-$7^2D$ transition in Cs and $5^2S$-$6^2D$ two-photon transition in Rb) that saturates the detector.

Figure 3: Computed RbCs potential energy curves (PEC) and transition dipole moments (TDM). The origin of energies is taken at the Rb($5S_{1/2}$)+Cs($6S_{1/2}$) limit. (a) The region around the Rb($5S_{1/2}$)+Cs($6P_{1/2,3/2}$) and Cs($6S_{1/2}$)+Rb($5P_{1/2,3/2}$) dissociation limits. Molecular states are denoted by their dominant Hund's case *a* character at short distances, with their Hund's case *c* symmetry reported as subscripts. The $(2)^3\Pi_{(\Omega=2)}$ curve is not drawn here for clarity sake. The horizontal rectangle shows the 30 cm$^{-1}$ range scanned with the PA laser around the Cs D2 line. The vertical dotted line indicates the location of the inner turning point of the a$^3\Sigma^+$ curve and of the minimum of the X$^1\Sigma^+$ ground state shown in panel (b). (c) TDMs for the four electric dipole-allowed transitions from the X and a states toward the four Hund's case *a* states $(3)^1\Sigma^+$, $(3)^3\Sigma^+$, $(2)^1\Pi$, $(2)^3\Pi$ correlated to the $Rb(5^2P)+Cs(6^2S)$ manifold.

Figure 4: Rotational constants of the vibrational levels of the $\Omega=0^+$, $0^-$, 1 states correlated to the $Rb(5P_{1/2,3/2})+Cs(6S_{1/2})$ manifold, computed within the coupled-state picture (see text). The rotational constants of Hund's case *a* states $(3)^1\Sigma^+$, $(3)^3\Sigma^+$, $(2)^1\Pi$, $(2)^3\Pi$ are also displayed for comparison. The vertical box illustrates the range explored with the PA laser. The origin of energies is taken at the $Rb(5S_{1/2})+Cs(6S_{1/2})$ limit.

Figure 5: Same as Figure 4 zoomed in to the region of the observed line (green squares with error bar) for the $\Omega=0^+$ (panel a), $0^-$ (panel c), 1 (panel e) states. In the other panels are drawn the weight on the Hund's case *a* electronic components of the vibrational levels computed in the coupled-state picture: (b) for $\Omega=0^+$, weight on the $(2)^3\Pi$ (full bar) and on the $(3)^1\Sigma^+$ (hashed bar); (d) for $\Omega=0^-$, weight on the $(2)^3\Pi$ (full bar) and on the $(3)^3\Sigma^+$ (hashed bar); for $\Omega=1$, weight on the $(2)^3\Pi$ (full bar), on the $(2)^1\Pi$ (hashed bar), and on the $(3)^1\Sigma^+$ (cross-hashed bar). Crosses locate the computed levels used in Figure 7 at transition energies 11715.1, 11716.9, 11722.4 cm$^{-1}$ for the $\Omega=0^+$, $0^-$, 1 states respectively. The vertical dotted line denotes the position of the $Rb(5S_{1/2})+Cs(6P_{3/2})$ limit.

Figure 6: Squared overlap integrals (closed circles) and Franck-Condon factors (open circles) between the initial radial wave function in the $a^3\Sigma^+$ state of the colliding $Rb(5^2S)$ and $Cs(6^2S)$ atoms at 290µK, and the weighted radial wave functions on the electronic components of the bound levels computed in the coupled-state picture: on the $(2)^3\Pi$ component of the $\Omega=0^+$ (a), $\Omega=0^-$ (b), $\Omega=1$ (d), levels, and on the $(3)^3\Sigma^+$ component of the $\Omega=0^-$ (c), $\Omega=1$ (e), levels. The vertical box illustrates the range scanned by the PA laser. The origin of energies is taken at the $Rb(5S_{1/2})+Cs(6S_{1/2})$ limit.

Figure 7: Squared overlap integrals between the radial wave functions of the $X^1\Sigma^+$ or $a^3\Sigma^+$ levels reached by spontaneous decay and the weighted radial wave functions on the electronic components of the PA bound levels (marked with crosses in Figure 5) computed in the coupled-state picture: the $a^3\Sigma^+$ levels wave functions with the $(2)^3\Pi$ component of the levels of (a) the $\Omega=0^+$, (b) $\Omega=0^-$, (c) $\Omega=1$, and the $X^1\Sigma^+$ levels wave functions with (d) the $(3)^1\Sigma^+$ component of the $\Omega=0^+$ levels, and with (e) the $(2)^1\Pi$ component of the $\Omega=1$ levels.

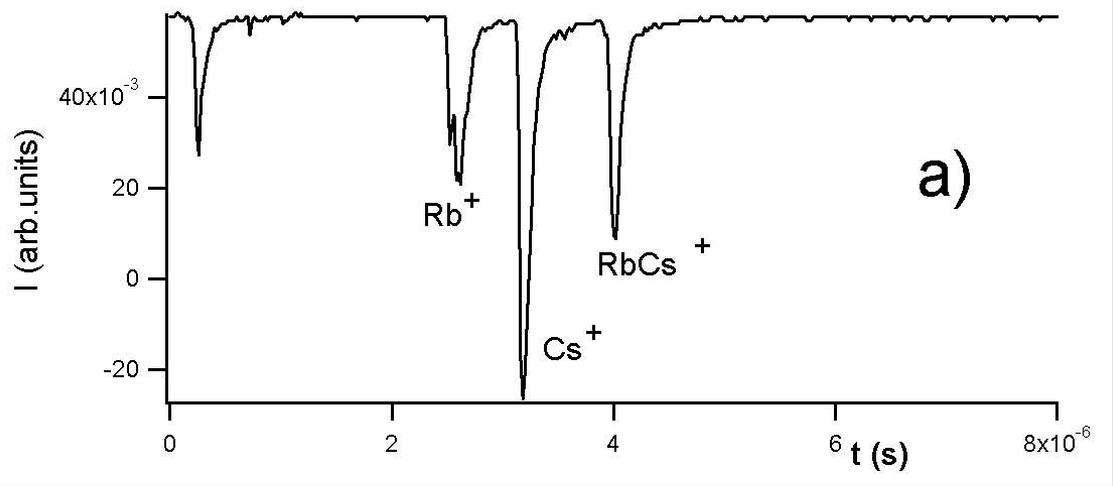
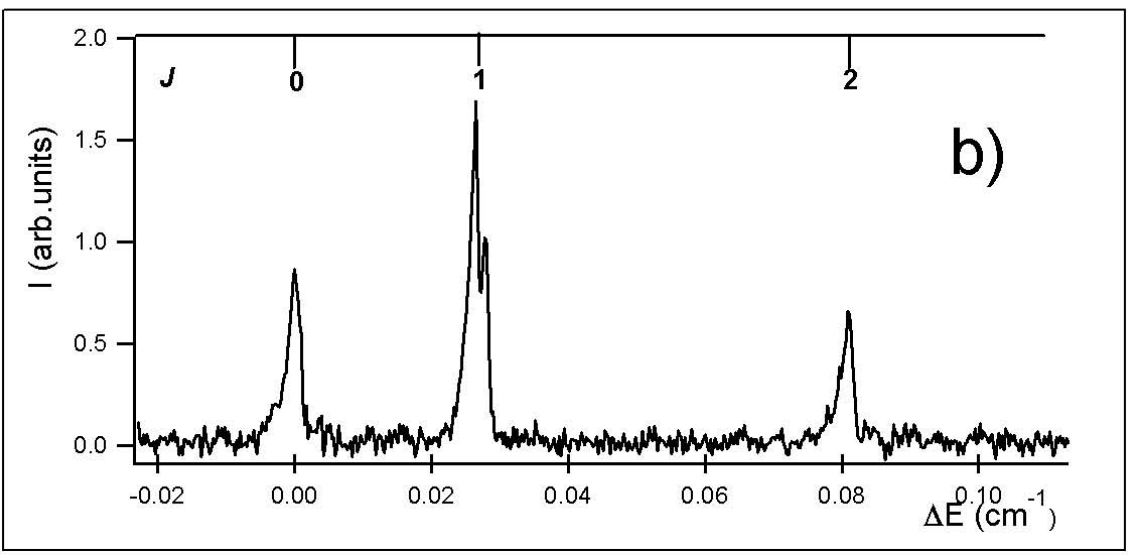
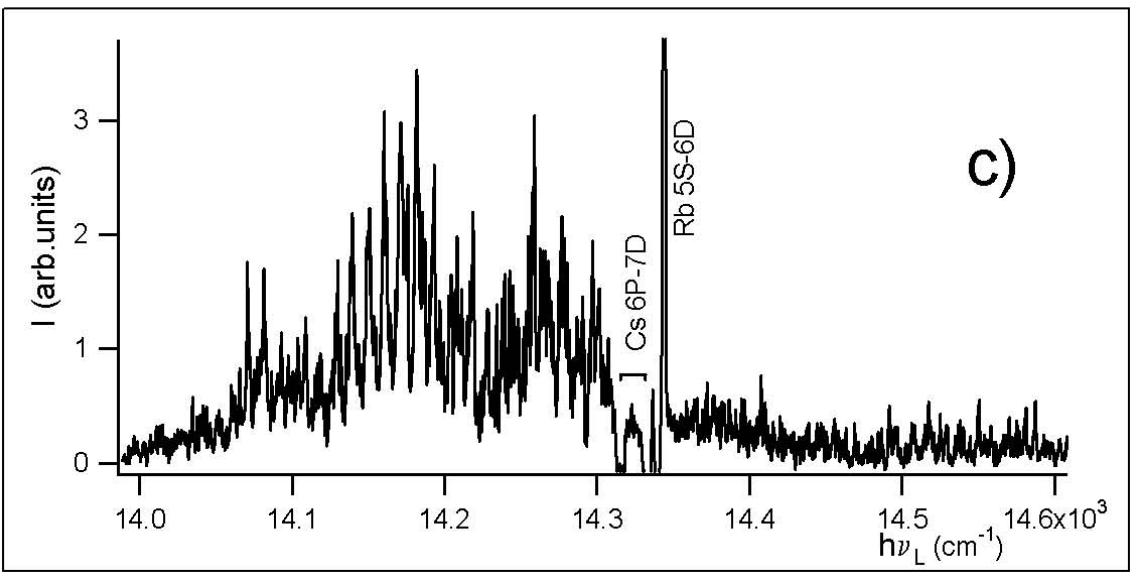

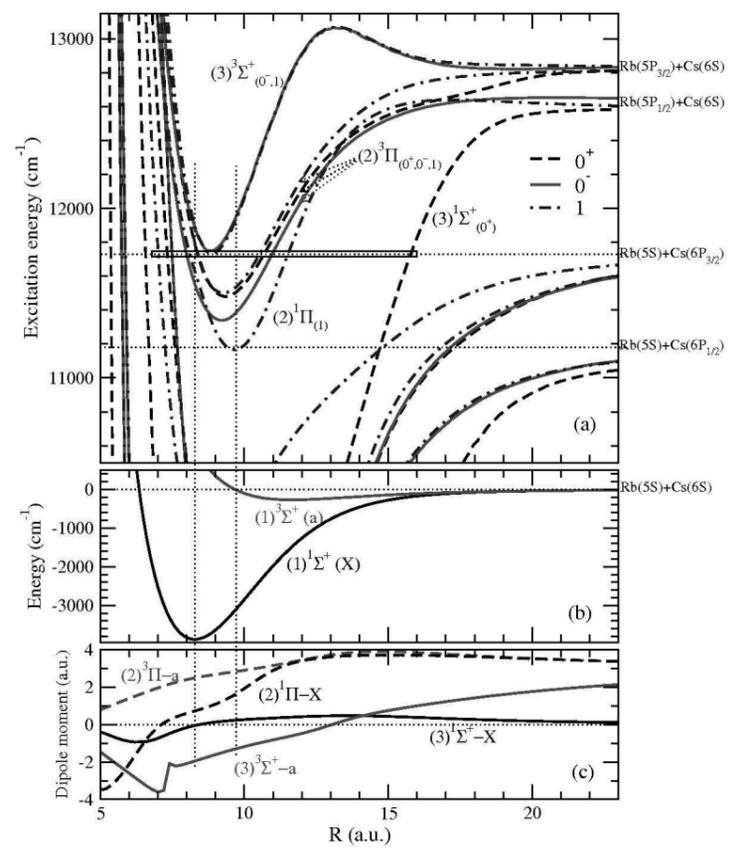

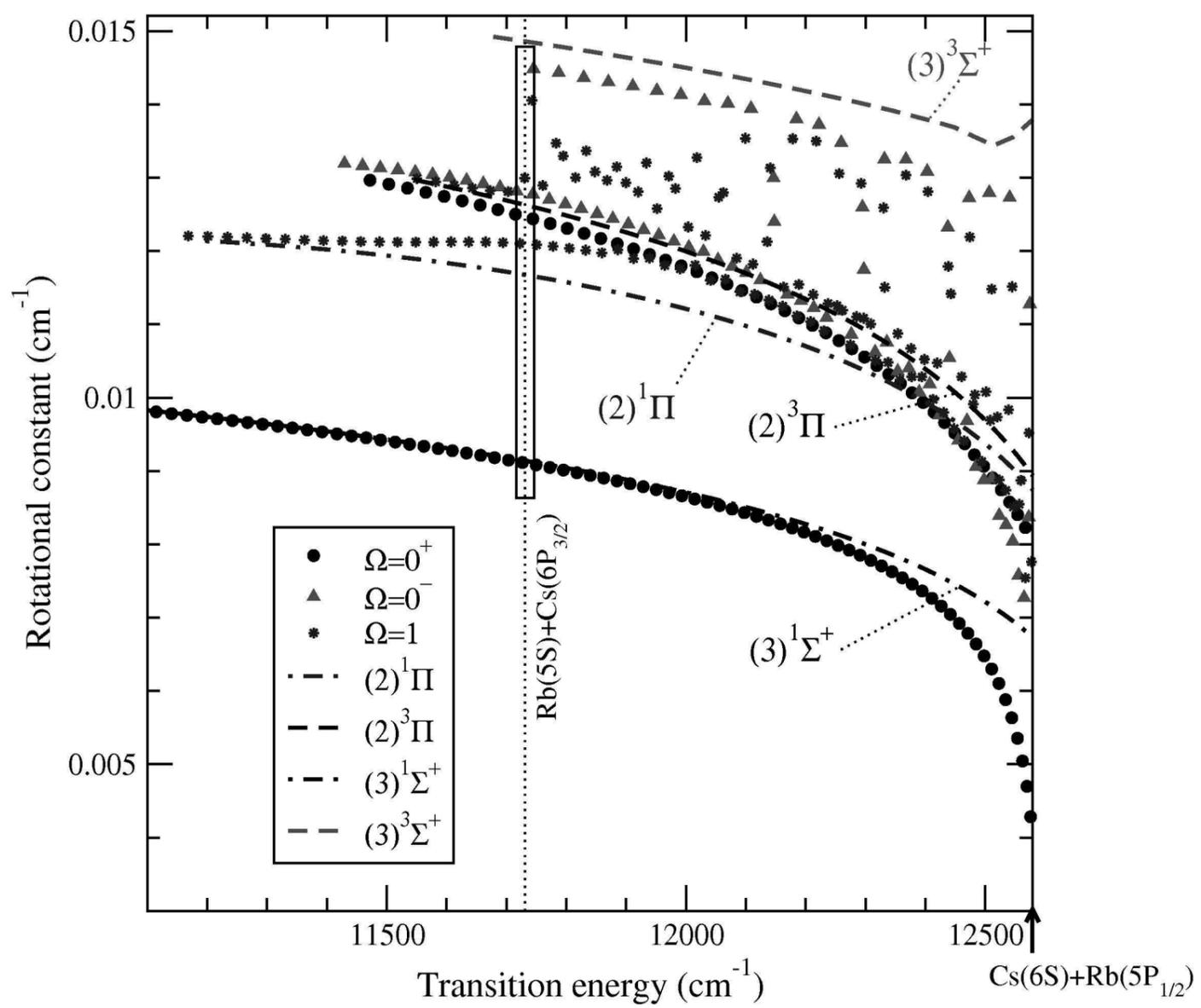

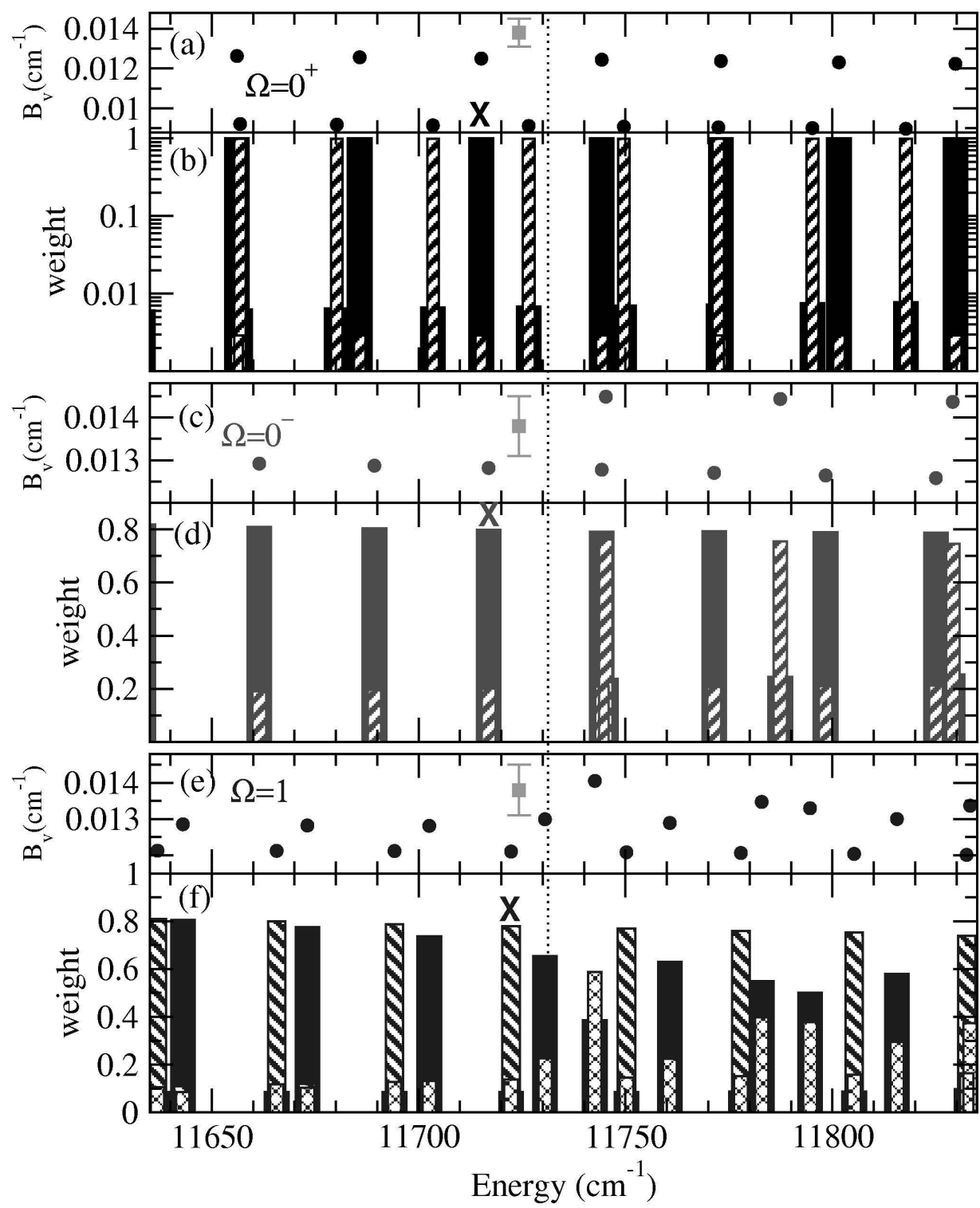

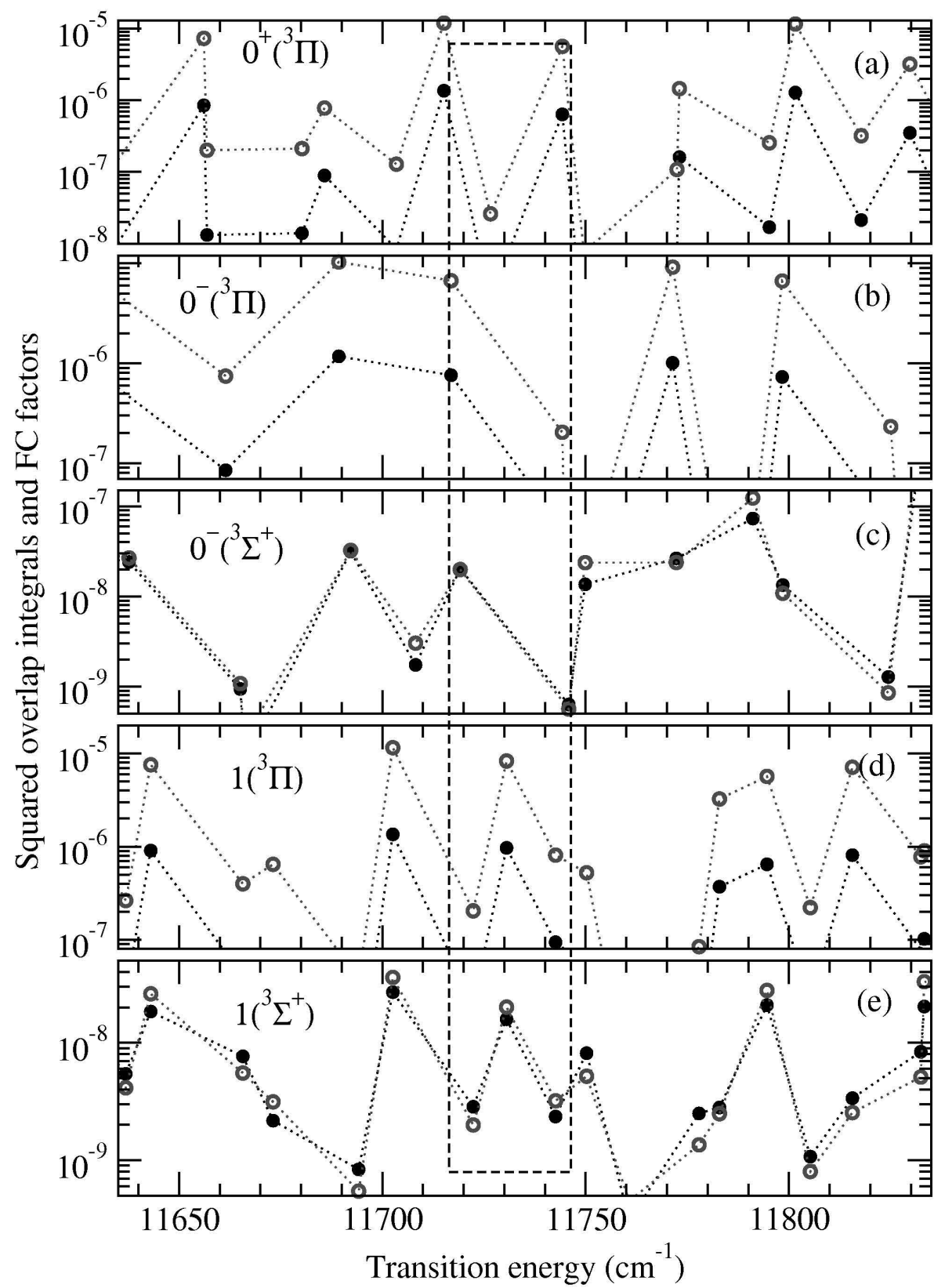

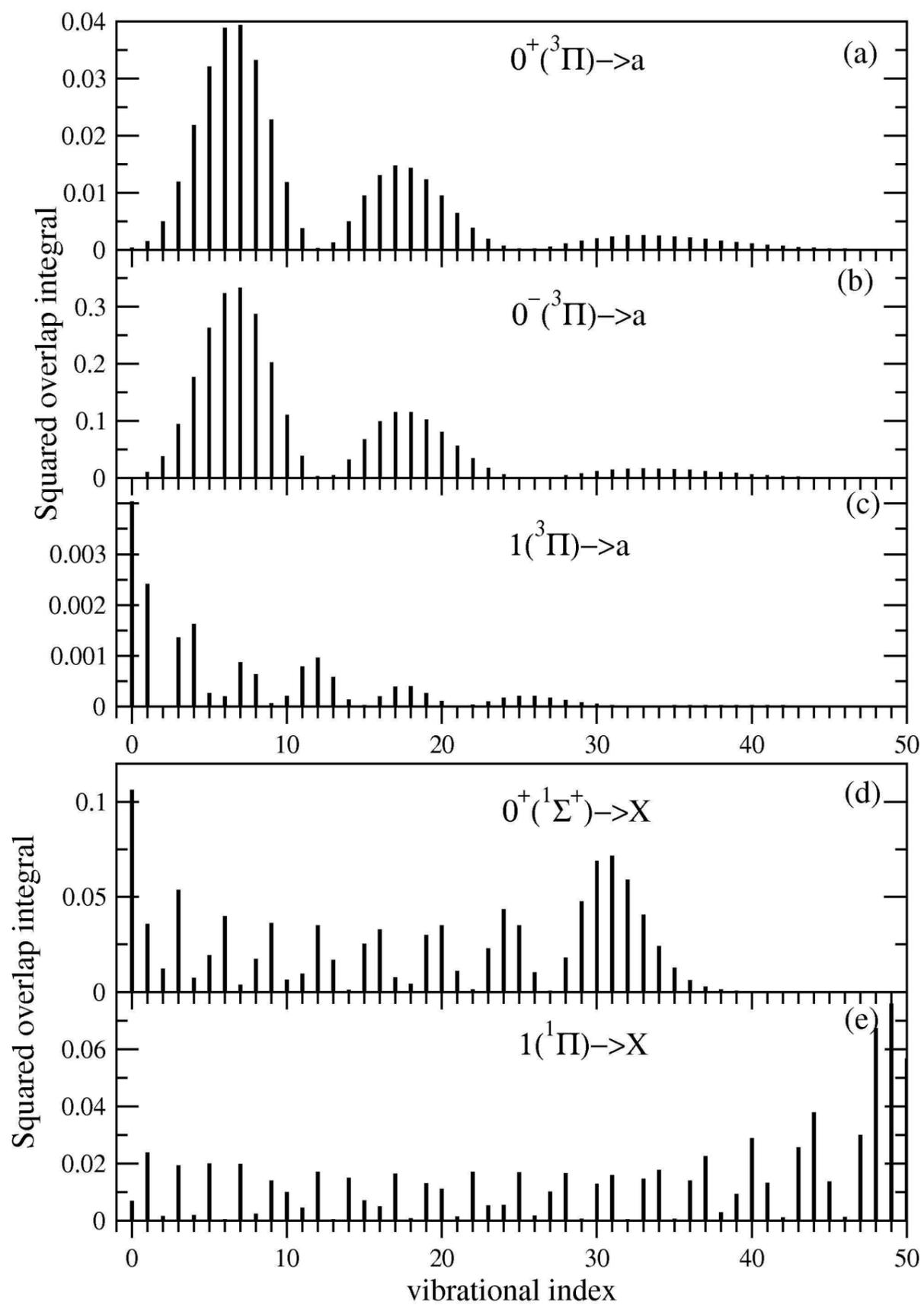